\documentstyle[aps,prl,preprint]{revtex}
\newcommand{\newc}{\newcommand}
\newc{\gsim}{\lower.7ex\hbox{$\;\stackrel{\textstyle>}{\sim}\;$}}
\newc{\lsim}{\lower.7ex\hbox{$\;\stackrel{\textstyle<}{\sim}\;$}}
\begin{document}
\begin{titlepage}

\baselineskip=14pt
\begin{flushright}
{\footnotesize
FERMILAB--Pub--96/213-A\\
UMD-PP-97-17\\
hep-ph/9608276\\
Submitted to Phys. Lett. {\bf B}}
\end{flushright}
\renewcommand{\thefootnote}{\fnsymbol{footnote}}
\vspace{0.15in}
\baselineskip=24pt

\begin{center}
{\Large \bf Higgs Boson Mass in Models with\\ Gauge-Mediated Supersymmetry
Breaking }\\
\baselineskip=14pt
\vspace{0.75cm}

\vspace{0.3cm}
{\bf A. Riotto$^{(1)}$\footnote{Electronic address:
{\tt riotto@fnal.gov}}, O. T\"{o}rnkvist$^{(1)}$\footnote{Electronic
address: {\tt olat@fnal.gov}} and
R.N. Mohapatra$^{(2)}$\footnote{Electronic
address: {\tt rmohapatra@umdhep.umd.edu}}}, \\
\vspace{0.4cm}
$^{(1)}${\em NASA/Fermilab Astrophysics Center,\\
Fermi National Accelerator Laboratory,\\ Batavia, Illinois~60510-0500, USA}\\
\vspace{0.3cm}
$^{(2)}${\em Department of Physics and Astronomy,
University of Maryland,\\ College Park, MD~~20742, USA}\\
\vspace{0.3cm}

\end{center}

\baselineskip=24pt

\begin{quote}
\hspace*{1em}
\begin{center}
{\bf\large Abstract}
\end{center}
\vspace{0.2cm}

We present the predictions for the mass $M_h$
of the lightest Higgs boson in
models with gauge-mediated supersymmetry breaking as a function
of the SUSY-breaking scale. We include all radiative corrections up to
two loops and point out that if the
CDF $e^+e^-\gamma\gamma$ event is interpreted in terms of these models,
then the lightest Higgs boson should be lighter than 110 GeV.

\vspace*{8pt}


\renewcommand{\thefootnote}{\arabic{footnote}}
\addtocounter{footnote}{-2}
\end{quote}
\end{titlepage}

\newpage
\def\NPB#1#2#3{Nucl. Phys. {\bf B#1} (19#2) #3}
\def\PLB#1#2#3{Phys. Lett. {\bf B#1} (19#2) #3}
\def\PLBold#1#2#3{Phys. Lett. {\bf#1B} (19#2) #3}
\def\PRD#1#2#3{Phys. Rev. {\bf D#1} (19#2) #3}
\def\PRL#1#2#3{Phys. Rev. Lett. {\bf#1} (19#2) #3}
\def\PRT#1#2#3{Phys. Rep. {\bf#1} (19#2) #3}
\def\ARAA#1#2#3{Ann. Rev. Astron. Astrophys. {\bf#1} (19#2) #3}
\def\ARNP#1#2#3{Ann. Rev. Nucl. Part. Sci. {\bf#1} (19#2) #3}
\def\MPL#1#2#3{Mod. Phys. Lett. {\bf #1} (19#2) #3}
\def\ZPC#1#2#3{Zeit. f\"ur Physik {\bf C#1} (19#2) #3}
\def\APJ#1#2#3{Ap. J. {\bf #1} (19#2) #3}
\def\AP#1#2#3{{Ann. Phys. } {\bf #1} (19#2) #3}
\def\RMP#1#2#3{{Rev. Mod. Phys. } {\bf #1} (19#2) #3}
\def\CMP#1#2#3{{Comm. Math. Phys. } {\bf #1} (19#2) #3}

\def\LHS{{\sc lhs}}
\def\RHS{{\sc rhs}}
\def\GUT{{\sc gut}}
\def\LTE{{\sc lte}}
\def\VEV{{\sc vev}}
\def\beq{\begin{equation}}
\def\eeq{\end{equation}}
\def\beqa{\begin{eqnarray}}
\def\eeqa{\end{eqnarray}}
\def\tr{{\rm tr}}
\def\ph{\widetilde{\gamma}}
\def\g{\widetilde{g}}
\def\x{{\bf x}}
\def\p{{\bf p}}
\def\k{{\bf k}}
\def\z{{\bf z}}
\def\re#1{{[\ref{#1}]}}
\def\eqr#1{{Eq.\ (\ref{#1})}}
\def\trh{{T_{\rm RH}}}
\def\tph{{T_{\rm PH}}}

\baselineskip=24pt
\renewcommand{\baselinestretch}{1.5}
\footnotesep=14pt


At present, supersymmetry (SUSY) is widely regarded as a leading
candidate for physics beyond the Standard Model \re{review}.
Although this is largely due
to the fact that SUSY provides the only known perturbative solution to the
problem of
quadratic divergence in the standard-model Higgs mass, its
additional virtues, such as providing a radiative mechanism to explain
the origin of the electroweak symmetry breaking and opening up possible
ways to unify gravity with other forces (via supergravity and
superstrings),
have made it especially appealing. Supersymmetry must of course
be a broken symmetry in order to agree with observations,
and an important unsolved problem of supersymmetric models is the nature
and the scale of SUSY-breaking. The most convenient approach is
to implement supersymmetry breaking in a hidden
sector and then transmit it to the standard-model sector in one of the
following two ways: either SUSY-breaking in the hidden sector
is conveyed to the
observable sector by
gravitational interactions; this is the so-called  $N=1$ supergravity
scenario \re{review}, or it is transmitted
via the gauge interactions of a distinct messenger sector \re{gmsb}
which contains fields that transform nontrivially under
the  standard-model gauge group.
In this paper we will be concerned with the latter class of models,
those with gauge-mediated supersymmetry breaking
(GMSB) \re{fc1}.

The effective low-energy theory that emerges from either of these
models contains soft SUSY-breaking mass terms for the scalar superpartners
which carry information about the scale and nature of the hidden-sector
theory. For instance, typical soft breaking terms for sfermions
resulting from the $N=1$ supergravity mechanism have magnitude
$\widetilde{m}^2\sim |F|^2/M_{{\rm Pl}}^2$, where $|F|$ is the
vacuum expectation value (VEV) of the $F$-term that breaks supersymmetry
in the hidden sector. In order to generate soft masses of the order of
$M_W$ in the matter sector, $\sqrt{|F|}$ should be around $10^{11}$ GeV.
On the other hand, in the  GMSB models where SUSY is broken at the scale
$\Lambda$, the magnitude of these terms
is given by  $\widetilde{m}^2\sim {{\alpha}\over{4\pi}}\Lambda$;
therefore, the same arguments imply a scale $\Lambda\lsim 10^2$
TeV which is much lower. This has
the interesting consequence that flavour-changing-neutral-current (FCNC)
processes are naturally suppressed in agreement with
experimental bounds.
The reason for this suppression is that the gauge interactions
induce flavour-symmetric SUSY-breaking terms in the observable sector at
$\Lambda$ and, because this scale is small, only a slight asymmetry
is introduced by renormalization group extrapolation to low energies.
This is in contrast to the supergravity scenarios
where one generically needs to invoke additional flavor symmetries to
achieve the same goal.

Another prediction of the GMSB models that distinguishes them from
$N=1$ supergravity models is the existence of
an ultra-light gravitino, $\widetilde{G}$
(which is the Nambu-Goldstone fermion corresponding to spontaneous
SUSY breaking), with mass given by
$M_{\widetilde{G}}\sim \Lambda^2/M_{{\rm Pl}}=
10^{-2}\left(\Lambda/10\:{\rm TeV}\right)^2$ eV. It is therefore the lightest
super-particle (LSP).
The recent observation by the CDF collaboration of a single event
with a final state
containing hard $e^+ e^{-}\gamma\gamma$ and missing transverse energy
\re{parke} can indeed be given a straightforward
interpretation in the context of GMSB models as selectron pair production
in a $p\bar{p}$ collision with $\widetilde{e}\rightarrow e +\widetilde{\gamma}$
followed by $\widetilde{\gamma}\rightarrow \gamma +\widetilde{G}$, and
similarly
for $\bar{\widetilde{e}}$ \re{exp1}.

An attractive feature of the GMSB models is that they are highly predictive.
Indeed, at energies well below the scale $\Lambda$, the theory looks like
the usual minimal supersymmetric standard model (MSSM) with the
remarkable difference that
all the free parameters (about 100) of the low-energy
supersymmetric standard
model are predicted in terms of three parameters: the SUSY-breaking scale
$\Lambda$, the $\mu$-parameter of the $H_d H_u$ term in the superpotential,
and the soft bilinear mass term, $B$.
Soft scalar masses $\widetilde{m}$ and gaugino masses $M$ are induced
at the two-loop and one-loop level, respectively, when the messenger
sector is
integrated out, and their values at the SUSY-breaking scale $\Lambda$  depend
only on  $\Lambda$.
Moreover, the trilinear soft breaking term $A$ vanishes at $\Lambda$. This
predictive power has recently been exploited to make a number of testable
predictions for the model \re{exp3}.

In this brief note we make use of the predictive power of the GMSB
models to compute the mass $M_h$ of the lightest $CP$-even state $h$
present in the Higgs sector as a function of the few parameters of the
GMSB models. We use the two-loop
corrected Higgs-boson mass spectrum to find accurate upper
bounds on the mass of the lightest Higgs boson and, in particular,
to extract any possible piece of information on $M_h$ obtainable
by combining these results with indications gathered from
sources such as the
CDF $ee\gamma\gamma$ event.
 In view of planned Higgs searches at LEP2 and LHC \re{lep}, we believe that
this prediction for $M_h$ can provide an additional test of the important
idea of gauge-mediated supersymmetry breaking.

The minimal GMSB models are defined by three sectors: (i) a secluded
sector that breaks supersymmetry; (ii) a messenger sector that serves
to communicate the SUSY breaking to the standard model and (iii) the
SUSY standard model.
The minimal messenger sector consists of a single
${\bf 5}+\bar{{\bf 5}}$ of $SU(5)$
(to preserve gauge coupling constant unification),
{\it i.e.} color triplets, $q$ and $\bar{q}$,
and weak doublets $\ell$ and $\bar{\ell}$
with their interactions determined by the following superpotential:
\begin{equation}
W=\lambda_1 S\bar{q}q+\lambda_2 S\bar{\ell}\ell.
\end{equation}
When the field $S$ acquires a VEV for both its scalar and auxiliary components,
$\langle S\rangle$ and $\langle F_S\rangle$ respectively,
the spectrum for $(q,\ell)$ is rendered non-supersymmetric.
Integrating out the messenger sector gives rise to gaugino
masses at one loop and scalar masses at two loops. For gauginos, we have
\begin{equation}
M_j(\Lambda)=k_j\frac{\alpha_j(\Lambda)}{4\pi}\Lambda,\:\:\:j=1,2,3,
\end{equation}
where $\Lambda= \langle F_S\rangle/\langle S\rangle$,
$k_1=5/3$, $k_2=k_3=1$ and $\alpha_1=\alpha/\cos^2\theta_W$. For the scalar
masses one has
\begin{equation}
\widetilde{m}^2(\Lambda)=2 \sum_{j=1}^3\:C_j
k_j\left[\frac{\alpha_j(\Lambda)}{4\pi}
\right]^2\Lambda^2,
\end{equation}
where $C_3=4/3$ for color triplets, $C_2=3/4$ for weak doublets
(and equal to zero otherwise) and $C_1=Y^2$ with $Y=Q-T_3$.
Because the scalar masses are functions of only the gauge quantum numbers,
these models automatically solve the supersymmetric flavor problem.
Notice the structure of the theory at this level. Squarks are the most
massive fields, their masses being roughly a factor of three higher than
the slepton masses.

These relations receive significant corrections from the renormalization
group evolution (RGE) from the scale $\Lambda$ down to the weak scale.
We have numerically solved the system of one-loop renormalization
group equations.
Radiative corrections drive the soft breaking mass squared
$m_{H_u}^2$ of the $H_u$-doublet, which couples to the top-quark,
to negative values near $M_Z$ leading to electroweak symmetry breaking.
They also raise slightly the soft breaking mass squared for
the sleptons. After including the effects of the RGE and $D$-terms,
the experimental limits on the right-handed selectron mass requires
\begin{equation}
\Lambda\gsim 10\:{\rm TeV}.
\end{equation}
We notice here that, if the $e^+ e^{-}\gamma\gamma$ plus
missing-transverse-energy event originates from slepton
pair-production
({\it e.g.} $\widetilde{e}_L\overline{\widetilde{e}}_L$ or
$\widetilde{e}_R\overline{\widetilde{e}}_R$), this restricts the
values of slepton
masses to  $(130\gsim m_{{\textstyle\widetilde{e}}_{L,R}}\gsim 80)$ GeV
\re{exp1}.
The $\widetilde{e}_R$-mode in turn implies that
\begin{equation}
(30\lsim \Lambda\lsim 50)\:{\rm TeV},
\end{equation}
whereas the $\widetilde{e}_L$-mode implies $20\lsim \Lambda \lsim 35$ TeV.
These upper bounds on $\Lambda$ will be used in the following to
constrain the mass of the lightest Higgs boson from above.

It is important to point out that
the magnitude of the $\mu$- and $B$-parameters at the scale $\Lambda$
depends crucially on the structure of the Higgs sector.
In the {\it minimal} messenger model, which contains only the usual
two Higgs doublets, one expects the $B$-parameter to be small at
the scale $\Lambda$ and to evolve to significant values at the
scale $M_Z$ in the process of running. In general,
in order to generate the parameters $\mu$ and $B$
at the scale $\Lambda$, the Higgs sector should be
enlarged \re{mu1},\re{mu2}.
However, this is not expected to affect the results of
this paper since, in general, the extra Higgs fields are so
heavy that they decouple from the matter fields at low energy.

Let us now consider the low-energy spectrum of the GMSB models as
far as the Higgs sector is concerned. As just mentioned,
we assume that its particle content at low energies is exactly that
of the MSSM. However, there are additional restrictions coming from the
structure of the GMSB theories.
The one-loop effective Higgs potential may be expressed as the sum
of the tree-level potential plus a correction coming from the sum of
one-loop diagrams with external lines having zero momenta,
\begin{equation}
V_{{\rm 1-loop}}=V_{{\rm tree}}+\Delta V_1.
\end{equation}
The right-hand side is independent of the running scale $Q$ to one-loop
order.
The one-loop correction is given by (in the $\overline{DR}$-scheme)
\begin{equation}
\Delta V=\frac{1}{64 \pi^2}\sum_j(-1)^{2s_j}(2 s_j+1) m_j^4
\left({\rm ln}\frac{m_j^2}{Q^2}-\frac{3}{2}\right),
\end{equation}
where $m_j$ is the eigenvalue mass of the $j^{{\rm th}}$
particle with spin $s_j$ in the $(v_d,v_u)$ background, with
$v_d=\langle H_d^0\rangle$ and
 $v_u=\langle H_u^0\rangle$.
The tree-level part of the potential of the MSSM Higgs sector reads
\begin{eqnarray}
V_{{\rm tree}}&=& m_d^2|H_d|^2+m_u^2|H_u|^2-\left(m_3^2H_d H_u+
\:{\rm h.c.}\right)\nonumber\\
&+&\lambda_1|H_d|^4+\lambda_2|H_u|^4+\lambda_3|H_d|^2|H_u|^2
+\lambda_4|H_d H_u|^2.
\end{eqnarray}
Here
\begin{eqnarray}
\lambda_1&=&\lambda_2=\frac{g_1^2+g_2^2}{8},\nonumber\\
\lambda_3&=&\frac{g_2^2-g_1^2}{4},\nonumber\\
\lambda_4&=&-\frac{g_2^2}{2},
\end{eqnarray}
where $g_1$ and $g_2$ are the gauge couplings of the $U(1)_Y$
and $SU(2)_L$ gauge groups respectively, and
\begin{equation}
m_d^2=m_{H_d}^2+|\mu|^2, \:\:m_u^2=m_{H_u}^2+|\mu|^2,\:\:m_3^2=B\mu.
\end{equation}
The parameters of the potential are allowed to run;
that is, they vary with scale according to the RGE. We must
use the RGE to evolve the parameters of the potential to a
convenient scale such as $M_Z$ (where the experimental values of the gauge
couplings are determined). After the following redefinition
\begin{equation}
\overline{m}_i^2=m_i^2+\frac{\partial \Delta V}{\partial (v_i^2)},\:\:i=d,u,
\end{equation}
minimization of the potential yields the following conditions among
the parameters:
\begin{equation}
\label{con}
\frac{1}{2}M_Z^2=\frac{\overline{m}_d\:^2-\overline{m}_u^2{\rm tan}^2\beta}
{{\rm tan}^2\beta-1},
\end{equation}
\begin{equation}
B\mu=-\frac{1}{2}\left(\overline{m}_d^2+\overline{m}_u^2\right){\rm sin}2\beta,
\end{equation}
where ${\rm tan}\beta=v_u/v_d$.

After $M_Z^2$ has been fixed to its physical value, all masses may
be expressed in terms of only two parameters and we have chosen them
to be the SUSY-breaking scale $\Lambda$ and $\tan\beta$. The
$\mu$-parameter at the scale $M_Z$ is then fixed by \eqr{con}.

The minimization conditions lead to the determination of the
the tree-level mass $M_h^{{\rm tree}}=M_Z |\cos 2\beta|$ of the
lightest $CP$-even state  $h$ of the Higgs spectrum. However,
it is well-known that radiative corrections contribute significantly
to the physical mass $M_h$. The Higgs-boson mass was first
determined by the renormalization-group resummation of all-loop
leading log (LL) corrections in
\re{h1}. Some next-to-leading log (NTLL) corrections were further
introduced in \re{h5} and \re{h6}, and finally a complete NTLL analysis
was performed in \re{h7} and \re{h8}. One of the main issues in \re{h8}
was the comparison between the LL and the NTLL approximations. As expected, the
LL approximation shows a strong scale-dependence, while the NTLL is almost
scale-independent. This implies not only that, working in the NTLL
approximation, the choice of scale is almost irrelevant, but also that the
LL approximation may yield accurate results if a correct choice of the
renormalization scale is made. The scale where both results coincide
turns out
to be close to the pole top-quark mass $M_t$ \re{h8}.

Very useful analytical approximations to the numerical all-loop
renormalization-group improved LL result, including two-loop
leading-log effects, may be found in \re{h9} where the reader
is referred to for more details. We report here the expression
for $M_h$ only in the case in which the mass $M_A$ of the $CP$-odd state in the
Higgs spectrum is much larger than $M_Z$\footnote{In this case all degrees
of freedom except the lightest $CP$-even state decouple,
leaving an effective
theory which is similar to the standard model with different
boundary conditions for the Higgs quartic coupling. In the opposite
case $M_A\lsim M_Z$, $M_h$ depends on $M_A$, see \re{h9}.}:
\begin{eqnarray}
\label{mh}
M_h^2&=&M_Z^2\cos^22\beta\left(1-\frac{3}{8\pi^2}\frac{m_t^2}{v^2}t\right)
\nonumber\\
&+&\frac{3}{4 \pi^2}\frac{m_t^4}{v^2}\left[\frac{1}{2}\widetilde{X}_t+t+
\frac{1}{16\pi^2}\left(\frac{3}{2}\frac{m_t^2}{v^2}-32\pi\alpha_3\right)
\left(\widetilde{X}_t t+t^2\right)\right],
\end{eqnarray}
where $v^2=v_d^2+ v_u^2$,
\begin{equation}
t={\rm ln}\left(\frac{M_{{\rm S}}^2}{M_t^2}\right),
\end{equation}
\begin{equation}
m_t=\frac{M_t}{1+\frac{4}{3\pi}\alpha_3(M_t)}
\end{equation}
is the on-shell running mass and $\alpha_3$ indicates
\begin{equation}
\alpha_3(M_t)=\frac{\alpha_3(M_Z)}{1+\frac{b_3}{4\pi}\alpha_3(M_Z)
{\rm ln}(M_t^2/M_Z^2)},
\end{equation}
where $b_3$ is the one-loop QCD beta function. Moreover,
$\widetilde{X}_t$ is the stop mixing parameter
\begin{eqnarray}
\widetilde{X}_t&=&\frac{2\widetilde{A}_t^2}{M_{{\rm S}}^2}\left(1-
\frac{\widetilde{A}_t^2}{12\:M_{{\rm S}}^2}\right),\nonumber\\
\widetilde{A}_t&=&A_t-\mu\:{\rm cot}\beta.
\end{eqnarray}
The expressions above assume that only the squarks of the third
generation contribute to the radiative corrections (this translates into
the bound $\tan\beta\lsim 35$).

The scale $M_{{\rm S}}$ is to be associated with the characteristic
stop mass scale and we have computed it in the following way. We have
solved the RGE for the soft SUSY-breaking parameters which enter the
stop mass matrix. They are  $\widetilde{A}_t$, $\widetilde{m}_Q$ and
$\widetilde{m}_U$, where the latter are the soft SUSY-breaking mass
terms of the left-handed and right-handed stop, respectively. The
initial conditions at the scale
$\Lambda$ are given by Eq. (3) and by
\begin{equation}
A_t(\Lambda)=0.
\end{equation}
Defining the stop squared-mass eigenvalues by $M_{\widetilde{t}_1}$
and   $M_{\widetilde{t}_2}$, the scale $M_{{\rm S}}$ has been defined
as the scale at which\footnote{Since, strictly speaking,  the operator
expansion leading to the expression (14) is performed in the symmetric phase,
one should have used the product of the SUSY-breaking squared masses
$\widetilde{m}_Q(M_{{\rm S}})\widetilde{m}_U(M_{{\rm S}})$ to define the scale
$M_{{\rm S}}$. We have checked that the numerical shift in the final result for
$M_h$ is negligible when adopting this definition instead of the one in Eq.
(\ref{msdef}).}
\begin{equation}
\label{msdef}
M_{\widetilde{t}_1}(M_{{\rm S}})M_{\widetilde{t}_2}(M_{{\rm S}})=
M_{{\rm S}}^2.
\end{equation}
Other operative definitions are possible, for example $M_{{\rm S}}^2=
(M_{\widetilde{t}_1}^2(M_{{\rm S}})+M_{\widetilde{t}_2}^2(M_{{\rm S}}))/2$,
but these different  distinctions have no significant  impact on the final
result for
$M_h$. We have generally found the $\mu$-parameter to be so large that the
pseudoscalar mass $M_A$ is driven to values much larger than $M_Z$,
rendering
the expression for $M_h$ in \eqr{mh} very reliable. Notice that the expression
(\ref{mh}), which we made use of in the case $M_A\gg M_Z$, is only valid
under the assumption (see Refs. \re{h9} for a thorough discussion)
\begin{equation}
\frac{M_{\widetilde{t}_1}^2(M_{{\rm S}})-M_{\widetilde{t}_2}^2(M_{{\rm S}})}
{M_{\widetilde{t}_1}^2(M_{{\rm S}})+M_{\widetilde{t}_2}^2(M_{{\rm S}})}\lsim
0.5.
\end{equation}
We have checked numerically that this condition was satisfied.

In Fig.~1 and Fig.~2, we present our predictions for
the mass of the lightest Higgs boson in the GMSB models as a function of
the scale $\Lambda$ for different values of $M_t$ and $\tan \beta$. From
Fig.~1, we see that the values of $M_h$ for a top quark mass
of 175 GeV range from 85 to 110 GeV for $\Lambda=50$ TeV. Fig.~2
shows the $\Lambda$ dependence of $M_h$ for $M_t=175$ GeV. The
requirement that the CDF $ee\gamma\gamma$ event is explained
by the GMSB scenario constrains the values of $M_h$ to lie on the left-hand
side of the vertical lines which show the upper bounds on $\Lambda$ coming
from
the constraint $m_{\widetilde{e}_R}\lsim 130$ GeV (long-dashed line) and
$m_{\widetilde{e}_L}\lsim 130$ GeV (dashed line). We infer that
$M_h\lsim 110$ GeV if $\widetilde{e}_R$ leads to the
CDF event and $M_h\lsim 105$ GeV for the $\widetilde{e}_L$ case.
Interestingly enough, this mass range  is accessible at LEP2 with a
center-of-mass energy  $\sqrt{s}=205$ GeV. This opens the exciting possibility
that one can obtain useful information about the GMSB models once these ranges
of Higgs masses are explored at LEP2 and LHC.

\vspace{36pt}

\centerline{\bf ACKNOWLEDGMENTS}

The work of A.R. is supported by the DOE and NASA under Grant NAG5--2788,
O.T. is supported in part
by the Swedish Natural Science Research Council (NFR),
and R.N.M. is supported by a grant from the National Science Foundation.


\def\NPB#1#2#3{Nucl. Phys. {\bf B#1} (19#2) #3}
\def\PLB#1#2#3{Phys. Lett. {\bf B#1} (19#2) #3}
\def\PLBold#1#2#3{Phys. Lett. {\bf#1B} (19#2) #3}
\def\PRD#1#2#3{Phys. Rev. {\bf D#1} (19#2) #3}
\def\PRL#1#2#3{Phys. Rev. Lett. {\bf#1} (19#2) #3}
\def\PRT#1#2#3{Phys. Rep. {\bf#1} (19#2) #3}
\def\ARAA#1#2#3{Ann. Rev. Astron. Astrophys. {\bf#1} (19#2) #3}
\def\ARNP#1#2#3{Ann. Rev. Nucl. Part. Sci. {\bf#1} (19#2) #3}
\def\MPL#1#2#3{Mod. Phys. Lett. {\bf #1} (19#2) #3}
\def\ZPC#1#2#3{Zeit. f\"ur Physik {\bf C#1} (19#2) #3}
\def\APJ#1#2#3{Ap. J. {\bf #1} (19#2) #3}
\def\AP#1#2#3{{Ann. Phys. } {\bf #1} (19#2) #3}
\def\RMP#1#2#3{{Rev. Mod. Phys. } {\bf #1} (19#2) #3}
\def\CMP#1#2#3{{Comm. Math. Phys. } {\bf #1} (19#2) #3}

\frenchspacing
\begin{picture}(400,50)(0,0)
\put (50,0){\line(350,0){300}}
\end{picture}

\vspace{0.25in}

\def\labelenumi{[\theenumi]}

\begin{enumerate}

\item\label{review} For a review, see, H.P. Nilles, Phys. Rep.
{\bf 110} (1984) 1; H.E. Haber and G.L. Kane, Phys. Rep. {\bf 117} (1985) 75;
A. Chamseddine, R. Arnowitt and P. Nath, {\it Applied N=1 Supergravity},
World Scientific, Singapore (1984).

\item\label{gmsb} M.~Dine, W.~Fischler, and M.~Srednicki, \NPB{189}{81}{575};
S.~Dimopoulos and S.~Raby, \NPB{192}{81}{353};
M.~Dine and W.~Fischler, \PLB{110}{82}{227};
M.~Dine and M.~Srednicki, \NPB{202}{82}{238};
M.~Dine and W.~Fischler, \NPB{204}{82}{346};
L.~Alvarez-Gaum\'e, M.~Claudson, and M.~Wise, \NPB{207}{82}{96};
C.R.~Nappi and B.A.~Ovrut, \PLB{113}{82}{175};
S.~Dimopoulos and S.~Raby, \NPB{219}{83}{479}.

\item\label{fc1} M.~Dine, A.E.~Nelson, and Y.~Shirman,
\PRD{51}{95}{1362};
M.~Dine, A.E.~Nelson, Y.~Nir, and Y.~Shirman, Phys. Rev. {\bf D53} (1996) 2658;
M. Dine, hep-ph/9607294.

\item\label{parke} S.~Parke, representing the CDF Collaboration,
{\it Search for New Phenomena in CDF}, in {\it 10th Topical Workshop
on Proton-Antiproton
	Collider Physics}, ed. by R.~Raha and J.~Yoh (AIP Press, New York,
	1995), FERMILAB-CONF-95/155-E.

\item\label{exp1} S.~Dimopoulos, M.~Dine, S.~Raby and S.~Thomas,
\PRL{76}{96}{3494};
S.~Ambrosanio, G.~Kane, G.~Kribs, S.~Martin and S.~Mrenna,
	\PRL{76}{96}{3498};  hep-ph/9607414.

\item\label{exp3}  K.S. Babu, C. Kolda
and F. Wilczek, IASSNS-HEP 96/55 preprint, hep-ph/9605408.

\item\label{lep} See, for instance, E. Accomando {\it et} al.,
{\it Higgs Physics}, hep-ph/9602250.



\item\label{mu1} G. Dvali, G.F. Giudice and A. Pomarol, CERN-TH/96-61
preprint, hep-ph/9603238.

\item\label{mu2} M. Dine, Y. Nir and Y. Shirman, SCIPP 96/30 preprint,
hep-ph/9607397.

\item\label{h1} R. Barbieri, M. Frigeni and F. Caravaglios, Phys. Lett. {\bf
B258} (1991) 167; Y. Okada, M. Yamaguchi and T. Yanagida, Phys. Lett. {\bf
B262} (1991) 54;
M. Carena, K. Sasaki and C.E.M. Wagner, Nucl. Phys. {\bf B381} (1992) 66;
P. Chankowski, S. Pokorski and J. Rosiek, Phys. Lett. {\bf B274} (1992) 191;
H.E. Haber and R. Hempfling, Phys. Rev. {\bf D48} (1993) 4280.

\item\label{h5} J.R. Espinosa and M. Quiros, Phys. Lett. {\bf B266} (1991) 389.

\item\label{h6} R. Hempfling and A.H. Hoang, Phys. Lett. {\bf B331} (1994) 99.

\item\label{h7} J. Kodaira, Y. Yasui and K. Sasaki, Phys. Rev. {\bf D50} (1994)
7035.

\item\label{h8} J.A. Casas, J.R. Espinosa, M. Quiros and A. Riotto, Nucl. Phys.
{\bf B436} (1995) 3.

\item\label{h9} M. Carena, J.R. Espinosa, M. Quiros and C.E.M. Wagner,
Phys. Lett. {\bf B355} (1995) 221; M. Carena,  M. Quiros and C.E.M. Wagner,
Nucl. Phys. {\bf B461} (1996) 407.

\end{enumerate}
\newpage
{\bf\large Figure Captions}
\vspace{1cm}

{\bf Fig. 1}: The physical mass of the lightest $CP$-even state $M_h$ as a
function of $M_t$ for $\tan\beta=2, 15$ and $\Lambda=50$ TeV.

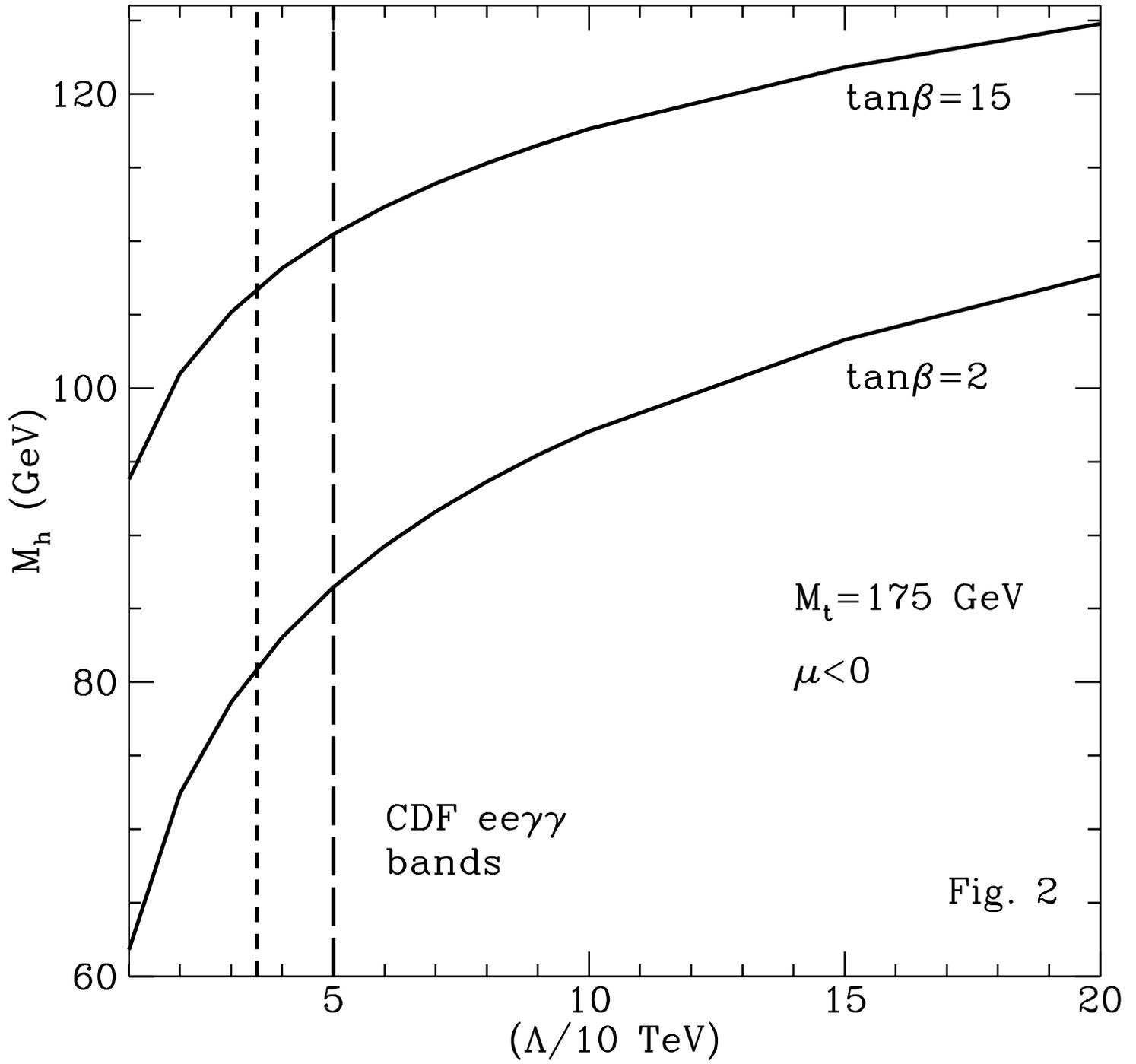
{\bf Fig. 2}: The physical mass of the lightest $CP$-even state $M_h$ as a
function of the SUSY-breaking scale $\Lambda$ for $\tan\beta=2, 15$ and
$M_t=175$ GeV. The dashed and  the long-dashed vertical lines
indicate the kinematical upper bounds on $\Lambda$ from the interpretation
of
the CDF $e e \gamma\gamma$ event as originating from $\widetilde{e}_L
\widetilde{e}_L$ and
$\widetilde{e}_R \widetilde{e}_R$ production, respectively.

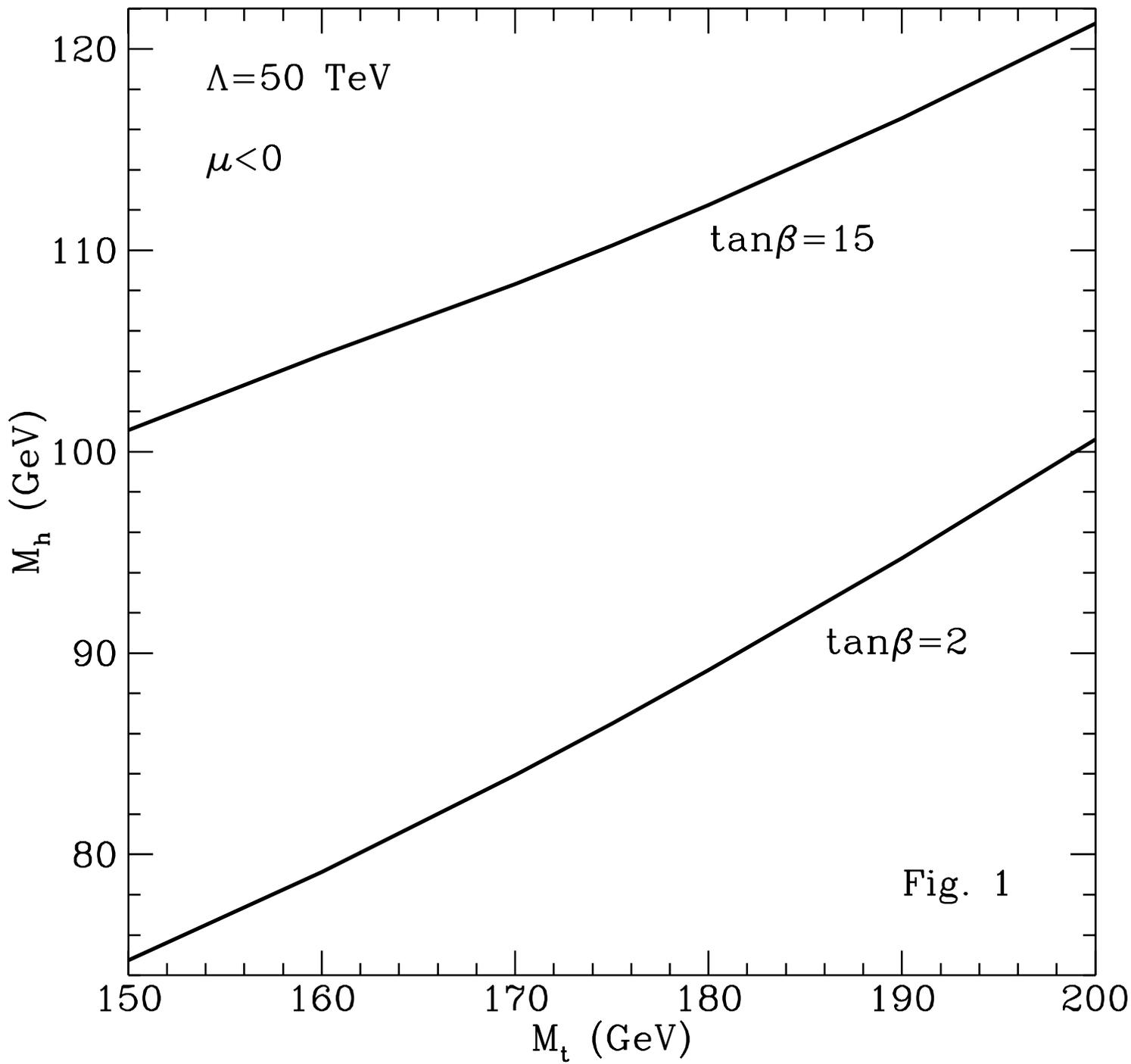
\end{document}